\documentclass[10pt,conference,letterpaper,nofonttune]{IEEEtran}

\usepackage{graphicx}
\usepackage{amsmath}
\usepackage{amsfonts}
\usepackage{optidef}
\usepackage{multirow}
\usepackage{xcolor}

\usepackage[font=footnotesize]{subcaption}
\usepackage[font=footnotesize]{caption}

\usepackage[draft]{hyperref}
\usepackage{comment}
\usepackage{lipsum}
\usepackage{graphicx}
\usepackage[utf8]{inputenc}
\usepackage{enumitem}
\usepackage{hyperref}
\usepackage{amsmath}
\usepackage{amsfonts}
\usepackage{bm}
\usepackage{tabularx}
\usepackage{svg}
\usepackage{algpseudocode}

\usepackage[linesnumbered,noend]{algorithm2e}

\AtBeginDocument{%
  \providecommand\BibTeX{{%
    \normalfont B\kern-0.5em{\scshape i\kern-0.25em b}\kern-0.8em\TeX}}}

\usepackage{xspace}
\SetKwRepeat{Do}{do}{while}

\usepackage[acronyms,nonumberlist,nopostdot,nomain,nogroupskip,hyperfirst=false]{glossaries}
\usepackage[square, numbers, sort&compress]{natbib}
\usepackage{balance}
\newacronym{iot}{IoT}{Internet of Things}
\newacronym{ss}{SS}{Spectrum Sensing}
\newacronym{dl}{DL}{deep learning}
\newacronym{dnn}{DNN}{Deep Neural Network}
\newacronym{cr}{CR}{Cognitive Radio}
\newacronym{cnn}{CNN}{Convolutional Neural Network}
\newacronym{lstm}{LSTM}{long-short term memory}
\newacronym{snr}{SNR}{signal-to-noise ratio}
\newacronym{stft}{STFT}{short-time fourier transform}
\newacronym{pu}{PU}{primary user}
\newacronym{sota}{SOTA}{state-of-the-art}
\newacronym{iq}{I/Q}{in-phase/quadrature}
\newacronym{iou}{IoU}{Intersection over Union}
\newacronym{ota}{OTA}{over-the-air}
\newacronym{rf}{RF}{Radio Frequency}
\newacronym{yolo}{YOLO}{You Only Look Once}
\newacronym{itw}{ITW}{in-the-wild}
\newacronym{ism}{ISM}{industrial, scientific and medical}
\newacronym{adc}{ADC}{analog-to-digital converter}
\newacronym{ai}{AI}{Artificial Intelligence}
\RestyleAlgo{ruled}

\usepackage[many]{tcolorbox}
\usepackage{lipsum}

\definecolor{titlebg}{RGB}{100,22,72}
\definecolor{introbg}{RGB}{0,128,128}

\newtcolorbox{usecase}[1][]{
  breakable,
  enhanced,
  arc=0pt,
  outer arc=0pt,
  colframe=titlebg,
  colback=titlebg!05,
  overlay unbroken and first={
    \node[
      draw=titlebg,
      fill=titlebg,
      rotate=0,
      anchor=north west,
      text=white,
      font=\bfseries
    ]
    at (frame.north west)  
    {#1};
  }
}

\newtcolorbox{mission}[1][]{
  breakable,
  enhanced,
  arc=0pt,
  outer arc=0pt,
  colframe=introbg,
  colback=introbg!05,
  overlay unbroken and first={
    \node[
      draw=introbg,
      fill=introbg,
      rotate=0,
      anchor=north west,
      text=white,
      font=\bfseries
    ]
    at (frame.north west)  
    {#1};
  }
}

\usepackage{pgfplots}
\usepackage{tikz}
\usepackage{pgfplotstable}
\usepackage[utf8]{inputenc}
\pgfplotsset{compat=1.18}
\newlength\fheight
\newlength\fwidth
\usetikzlibrary{plotmarks,patterns,decorations.pathreplacing,backgrounds,calc,arrows,arrows.meta,spy,matrix}
\usepgfplotslibrary{patchplots,groupplots,colorbrewer}
\usepackage{tikzscale}
\usepackage{xfrac}

\usepackage{tikz}
\usepackage{tikzscale}
\newif\ifexttikz
\exttikzfalse
\ifexttikz
\else
\usepackage{tikzpagenodes,etoolbox}
\usetikzlibrary{calc}
\usepackage[contents={}]{background}
\AddEverypageHook{%
\ifnumequal{\thepage}{1}{%
    \tikz[remember picture,overlay]{%
        \node[draw,
        minimum width=1.03\textwidth,
        text width=1.02\textwidth,
        font=\footnotesize
        ]
        at ($(current page header area) - (0,5pt)$)
        {%
        This paper has been accepted for publication on IEEE International Conference on Computer Communications (INFOCOM). This is the author's accepted version of the article. The final version published by IEEE is D. Uvaydov, M. Zhang, C.P. Robinson, S. D'Oro, T. Melodia, F. Restuccia, "Stitching the Spectrum: Semantic Spectrum Segmentation with Wideband Signal Stitching," Proc. of IEEE International Conference on Computer Communications (INFOCOM), Vancouver, BC, Canada, May 2024.
        };
        \node[draw,
        minimum width=1.03\textwidth,
        text width=1.02\textwidth,
        font=\footnotesize
        ]
        at (current page footer area)
        {%
        ©2024 IEEE. Personal use of this material is permitted. Permission from IEEE must be obtained for all other uses, in any current or future media, including reprinting/republishing this material for advertising or promotional purposes, creating new collective works, for resale or redistribution to servers or lists, or reuse of any copyrighted component of this work in other works.
        };
    }%
}{}
}
\fi

\begin{document}

\title{\textit{Stitching the Spectrum:} Semantic Spectrum Segmentation with Wideband Signal Stitching\vspace{-0.3cm}}


\author{\IEEEauthorblockN{Daniel Uvaydov$^{*}$, Milin Zhang$^{*}$, Clifton Paul Robinson, \\Salvatore D'Oro, Tommaso Melodia and Francesco Restuccia\vspace{-0.3cm}}\\
\IEEEauthorblockA{
Institute for the Wireless Internet of Things, Northeastern University, United States\\ 
Email: {\{uvaydov.d, zhang.mil, robinson.c, s.doro, t.melodia, f.restuccia\}}@northeastern.edu}\vspace{-0.4cm}\\$*$ These authors contributed equally to this work \vspace{-0.5cm}
}

\maketitle

\begin{abstract}
Spectrum has become an extremely scarce and congested resource. As a consequence, spectrum sensing enables the coexistence of different wireless technologies in shared spectrum bands. Most existing work requires spectrograms to classify signals. Ultimately, this implies that images need to be continuously created from I/Q samples, thus creating unacceptable latency for real-time operations. In addition, spectrogram-based approaches do not achieve sufficient granularity level as they are based on object detection performed on pixels and are based on rectangular bounding boxes. For this reason, we propose a completely novel approach based on \textit{semantic spectrum segmentation}, where multiple signals are simultaneously classified and localized in both time and frequency at the I/Q level. Conversely from the state-of-the-art computer vision algorithm, we add non-local blocks to combine the spatial features of signals, and thus achieve better performance. In addition, we propose a novel data generation approach where a limited set of easy-to-collect real-world wireless signals are ``stitched together'' to generate large-scale, wideband, and diverse datasets. Experimental results obtained on multiple testbeds (including the Arena testbed) using multiple antennas, multiple sampling frequencies, and multiple radios over the course of 3 days show that our approach classifies and localizes signals with a mean intersection over union (IOU) of 96.70\% across 5 wireless protocols while performing in real-time with a latency of 2.6 ms. Moreover, we demonstrate that our approach based on non-local blocks achieves 7\% more accuracy when segmenting the most challenging signals with respect to the state-of-the-art U-Net algorithm. We will release our 17 GB dataset and code.  
\end{abstract}


\vspace{-0.1cm}
\section{Introduction}


The demand for mobile connectivity is increasing spectrum congestion to unprecedented levels \cite{SpectrumCrunch}, with the number of mobile devices expected to reach 64 billion by 2025 \cite{EricssonMobilityReport2021}, and global mobile data traffic to reach 288 exabytes per month by 2027 \cite{5GAmericas_WP}.  To improve spectrum usage, technologies such as \textit{spectrum sharing} will become fundamental components of next-generation wireless systems \cite{zhang2017survey}. A fundamental problem in spectrum sharing is determining in real time which spectrum bands are currently underutilized, also referred to as \emph{spectrum sensing} \cite{arjoune2019comprehensive}. Traditionally, spectrum sensing has focused on determining whether a given channel is busy or free \cite{jin2018privacy,qi2018channel,liu2019deep,qi2019low,chew2020spectrum,uvaydov2021deepsense}. This, however, does not allow to implement effective spectrum policies where wireless technologies have different access priorities  \cite{baldesi2022charm}. For this reason, a much more relevant -- though significantly more challenging -- problem is determining which wireless technology is utilizing a given portion of the spectrum over time and frequency, also known as \textit{spectrum classification} \cite{mao2018deep}. To this end, recent work has demonstrated that data-driven approaches based on \gls{dl} can be effective in addressing spectrum classification tasks \cite{baldesi2022charm,uvaydov2021deepsense}.

\begin{figure}[!h]
    \centering
    \includegraphics[width=\columnwidth]{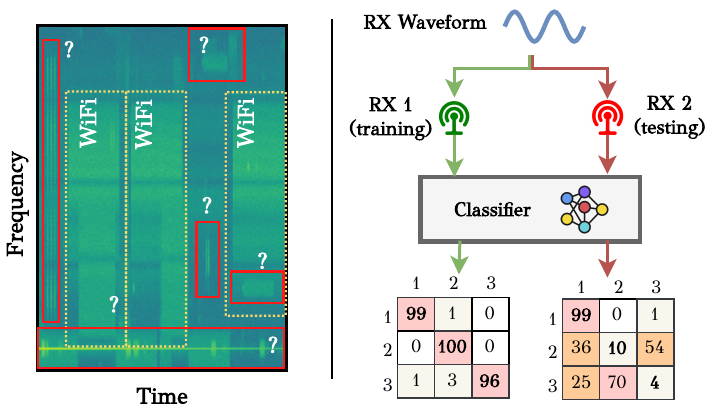}
    \caption{(Left): Example of why labeling spectrum is a hard problem; (Right): A waveform classifier may perform well on data collected by the same device used during the training phase (RX 1), but may show poor performance when tested on data collected by a different device (RX 2).} 
    \vspace{-0.2cm}
    \label{fig:dataset_lack} 
\end{figure}

Existing work on \gls{dl} for spectrum classification---discussed in Section \ref{sec:rw}---suffers from a number of critical issues. Most importantly, prior work has mostly relied on simulations and/or small-scale experimental datasets to evaluate performance \cite{8302156, 8357902, 9580446, vagollari2021joint, soltani2022finding}, this is not without reason. Indeed, labeling real-world wideband spectrum is extremely challenging due to the coexistence of different signals in the same spectrum bands. Figure~\ref{fig:dataset_lack} (left) shows an example of a spectrum capture obtained through our testbed---described in Section \ref{sec:experiment}---in the 5 GHz \gls{ism} band. Although it is relatively easy to properly label Wi-Fi transmissions, the collected data also shows several interfering signals at different frequencies whose type and location is not available a priori, which makes their labeling hardly feasible. 

In addition, the lack of diverse datasets necessarily leads to poor performance under dynamic channel conditions \cite{restuccia2019deepradioid,doro2021can}. An illustrative example is shown in Figure \ref{fig:dataset_lack} (right), where a waveform classifier achieves very high accuracy when tested with data collected by one of the devices used to generate the training dataset (i.e., RX 1), but performs poorly when tested with data collected \gls{ota} by a new and unseen device (i.e., RX 2) whose data was never included in the training dataset. Although it has been demonstrated that feeding as much diverse data (in terms of channel, mobility, and traffic conditions) makes \gls{dl} solutions more reliable, how to collect such data is an open challenge \cite{shawabka2020exposing}.

\begin{figure}[!t]
\includegraphics[width=\columnwidth]{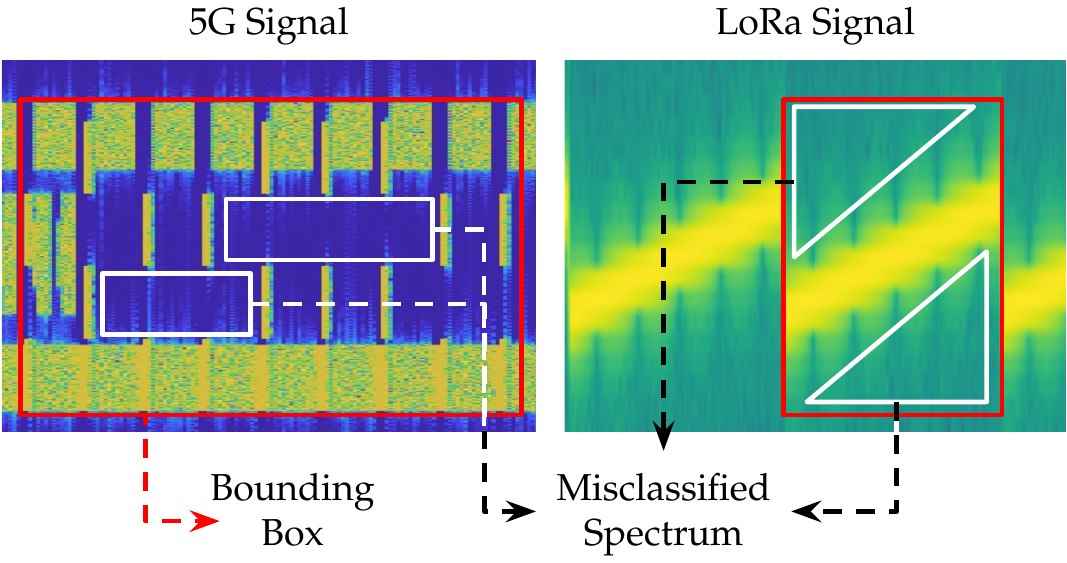}
\caption{Object detection requires the construction of an image and does not achieve fine-grained I/Q granularity.} 
\vspace{-0.2cm}
\label{fig:od-vs-seg}
\end{figure}

Lastly, existing work mostly leverages object detection algorithms such as \gls{yolo}, which are used out-of-the-box from computer vision applications \cite{kayraklik2022application, nguyen2023wrist}. This approach has several drawbacks. First, it requires the creation of an image out of \gls{iq} samples, thus incurring additional latency. This is a critical issue since spectrum needs to be classified almost instantaneously to provide meaningful and actionable inference -- for example, to detect and fill a spectrum hole. Moreover, modern wireless signals such as 5G and LoRa can hardly conform into square boxes, as shown in Figure \ref{fig:od-vs-seg}. Thus a significant amount of spectrum will be incorrectly classified as occupied, thus leading to poor spectrum efficiency if opportunistic and/or multi-tier spectrum access policies are being implemented \cite{FederatedWireless}.

\subsection*{Summary of Novel Contributions}

In this paper, we address the above three issues by proposing the following novel contributions: \smallskip

\textbf{(1)} We present a novel dataset generation pipeline for wideband spectrum sensing applications, which allows us to generate large-scale datasets that (i) contain signals collected \gls{ota} and are affected by real-world channel conditions; (ii) can be completely labeled; and (iii) can be generated in a time-efficient manner. Our approach is based on ``stitching'' different signals together to create samples where signals are overlapping and affected by real-world noise and interference. We show that this pipeline can generate a virtually infinite amount of labeled data starting from only 17 GB of \gls{ota} data captured using only two radios in 3 hours and for 5 different wireless technologies (i.e., WiFi, LTE, BLE, LoRa, and ZigBee); \smallskip

\textbf{(2)} We develop a novel custom \gls{dl} algorithm for multi-label multi-class spectrum sensing based on \textit{semantic segmentation}. Although our algorithm is structured based on the widely known U-Net network for image segmentation \cite{ronneberger2015u}, it presents a few key differences that make our network perform better for spectrum segmentation. Conversely from existing spectrum sensing work, our approach (i) operates at the \gls{iq} level, instead of creating images; (ii) classifies each and every \gls{iq} sample incoming from the ADC, without creating any bounding boxes, thus increasing classification accuracy significantly; (iii) it only uses 1024 \gls{iq} samples as input, which leads to very low inference time. Moreover, we show that our network achieves 7\% more accuracy than U-Net in the most challenging protocols (Wi-Fi and LTE) while maintaining similarly low latency; \smallskip

\textbf{(3)} We demonstrate experimentally how the proposed semi-augmented dataset generation pipeline can make \gls{dl}-based solutions more robust against changing conditions and capable of generalizing across different scenarios and deployments. Specifically, we show how the pipeline makes the above wideband spectrum sensing algorithm more accurate and able to deliver high accuracy even when operating with data collected by different devices, sampling rates, antennas, and under previously unseen network conditions. Our results show that we localize and classify 5 wireless protocols with a mean \gls{iou} of 96.70\%. We also performed real-world experiments with GPUs and we show that our algorithm takes only \textbf{2.6}~ms of latency to process 100 MHz of spectrum.


\section{Background and Related Work}\label{sec:rw}

\gls{dl} has been identified as the ideal candidate to solve a variety of challenges in the wireless domain, where the RF environment is hard to model using closed forms due to its non-deterministic behavior \cite{jagannath2019machine,luong2019applications}. 

Since \gls{dl}-based solutions require large quantities of data to be effective, one approach that finds broad application in the literature is that of using simulators and emulators to generate large synthetic datasets aiming to capture the different behavior of waveforms transmitted \gls{ota} under diverse channel conditions~\cite{li2019survey}. Although entirely relying upon synthetic data is sufficient to validate complex \gls{dl} algorithms and demonstrate their potential in solving complex networking tasks, this approach does not necessarily transfer well to real-world applications where the input data consists of signals collected \gls{ota}~\cite{8302156, 8357902, 9580446, vagollari2021joint, soltani2022finding}. 

Indeed, it has been shown that utilizing signals collected via experimental data collection campaigns allows \gls{dl} models to better generalize and achieve higher accuracy~\cite{8267032, soltani2019real}. 
For example, \cite{8267032, soltani2019real} design and train a \gls{dl} classifier that uses \gls{ota} data to perform waveform classification. Despite demonstrating the benefits and importance of using \gls{ota} data, these works only produce a single-class classification outcome and focus on modulation recognition tasks only, thus not providing complete information on spectrum utilization and resulting in lower accuracy when handling interfering and coexisting signals. 

Some more recent work aims at localizing wireless signals in the spectrum using well-established computer vision approaches meant for object detection, such as \gls{yolo}~\cite{kayraklik2022application, nguyen2023wrist}. As mentioned before, while object detection via \gls{yolo} is indeed a viable method for computer vision tasks, this approach does not achieve the level of resolution or granularity required for wireless signals. Wireless signals do not necessarily fit neatly into square bounding boxes (i.e. chirp spread spectrum signals). Furthermore, there are wireless technologies whose signals do not fully utilize their allocated spectrum (i.e. LTE, 5G) and bounding boxes can not accurately depict the under-utilization of these technologies, as shown in Fig. \ref{fig:od-vs-seg}. This is especially crucial in applications of opportunistic spectrum sharing where underutilized portions of the spectrum can be taken advantage of to increase spectrum efficiency. 

Another line of work focuses on using data augmentation techniques to diversify and extend the scope of datasets collected \gls{ota}. Specifically, \cite{soltani2020more, huang2019data} develop data augmentation pipelines for wireless signal classification with a specific focus on single-label classification tasks, where the augmentation process is designed to emulate the effect of diverse channel effects and noise levels on collected signals. 
If compared to ours, the above works focus on single-label classification tasks and do not address the problem of locating and characterizing multiple potentially overlapping wireless signals in wideband applications.

The closest paper to ours is the very recent work from Nguyen \textit{et al.}~\cite{nguyen2023wrist}, where authors curate a \gls{ota} dataset that is extended via data augmentation techniques to address wideband spectrum sensing tasks. In \cite{nguyen2023wrist}, authors develop a spectrum sensing algorithm that uses \gls{yolo}-based object detection to identify and localize different waveforms in real-time and \gls{itw}. Despite the approach being very similar to ours at a high level, \cite{nguyen2023wrist} utilizes \gls{yolo} which produces extensive bounding boxes that might include either empty portions of spectrum or signals belonging to a different class. Moreover, despite the authors of \cite{nguyen2023wrist} demonstrating exceptional accuracy results, they do not evaluate the portability and generalization of the solution against different radios, sampling rates, and in environments where transmissions are generated by devices not being controlled by the authors.

\textbf{Our work separates itself from the existing literature in that:} (i) it introduces a semi-augmented data generation pipeline that has been designed to facilitate the generation of \gls{ota} wideband datasets including waveforms with diverse technologies, channel and noise conditions, sampling rates, transmission antennas, and center frequency; (ii) it presents a novel multi-class \gls{dl}-based semantic spectrum segmentation algorithm capable of delivering high-resolution spectrum sensing capabilities that go beyond inaccurate bounding boxes used in \gls{yolo}-based solutions, thus offering a more accurate classification and localization of signals and unused portions of the spectrum; and (iii) we validate our work following a purely \gls{ota} approach with signals collected \gls{itw} with multiple radios, sampling rates, center frequencies, and RF environments. 

\section{Semi-Augmented Dataset Generator} \label{sec:data_gen}


Accurate \gls{dl}-based spectrum sensing heavily relies on the availability of properly labeled and diverse datasets. However, how to build such datasets via \gls{ota} data collection is resource- and time-consuming. This problem is further exacerbated when the inference objective is not simply classification (e.g., recognizing the modulation of a signal, or determining the presence of a specific waveform), but requires a more fine-grained output. Among others, a relevant spectrum sensing task is \textit{multi-label multi-class spectrum sensing,} where the goal is to monitor a portion of the spectrum and identify all signals that are being transmitted, their bandwidth, technology, and center frequency.

The main challenge is that data must be collected \gls{ota} so as to capture realistic channel conditions and prepare the \gls{ai} to operate correctly in a wireless deployment with real radios. 
Additionally, nodes performing spectrum sensing might not always be aware of where signals are being transmitted, and their center frequency will rarely match that of the signals the sensing node is trying to detect, classify and locate. In Section \ref{sec:experiment}, we demonstrate how training on data that always assumes synchronization between sensing and transmission center frequencies results in poor accuracy. 


Undoubtedly, one could perform a large \gls{ota} data collection campaign. However, such a task would be extremely hard and time-consuming due to (i) the several many possible combinations to be covered in time and frequency, which exponentially grow in the case of wideband spectrum sensing; and (ii) the need for large portions of spectrum without any interference from external systems. Our goal is to mitigate the complexity via a combination of \gls{ota} data collection and data augmentation. 

\textbf{Approach in a nutshell.}~First, we create a \textit{signal bank} (Section \ref{sec:bank}), i.e., a dataset containing individual signals collected \gls{ota} in the absence of interference. This serves as a seed to create much larger and more diverse datasets. Then, our generator (Section \ref{sec:generator}) extracts multiple signals from the signal bank and combines them together (via time and frequency shift mechanisms) to create a semi-augmented ``stitched signal" portraying a more realistic \gls{ota} data capture. These signals will interfere with each other, be in randomized frequency locations, and will have already experienced the common channel effects (fading, multipath, etc.) during \gls{ota} collection.


\subsection{Signal Bank Generation} \label{sec:bank}

The signal bank consists of a labeled collection of individual signals collected \gls{ota} when (i) only one signal is transmitted at a time; (ii) the center frequency and bandwidth are known a-priori; and (iii) data collection is performed over a limited and small portion of the spectrum that is constantly monitored to ensure the lack of interference from external systems. 

\begin{figure}[t!]
    \centering
    \includegraphics[width=\columnwidth]{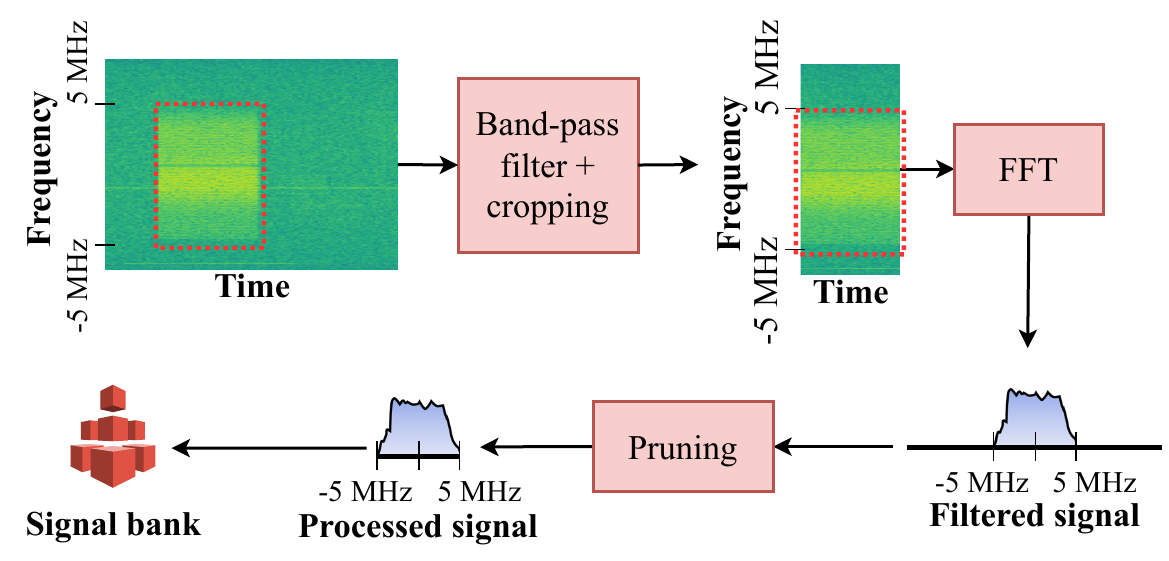}
    \caption{Pre-processing for frequency synchronized signals} 
    \label{fig:preproc} 
    \vspace{-0.5cm}
\end{figure}

To be added to the signal bank, signals first go through a pre-processing pipeline shown in Fig. \ref{fig:preproc}. Specifically, signals are broken up into shorter signals that are (i) cropped to contain only the actual signal transmission (e.g., by removing the silence period before and after data transmission); and (ii) bandpass-filtered to only extract the signal of interest and remove any undesired signals that may have been recorded. Once signals have been pre-processed, they are (iii) converted to the frequency domain through a Fast Fourier Transform (FFT). Finally, being in the frequency domain, (iv) any frequency components outside of the band of interest occupied by the signal are pruned, and the remaining \glspl{iq} are added to the signal bank. This procedure is repeated multiple times for each signal type so as to generate multiple instances of the same signal with diverse duration and spectrum occupancy. Note, we do not remove any small interference or channel effects that may overlap with the signal of interest during recording, we consider this to be commonplace effects that may happen on a signal and will add to the generalization ability later. The combination of signals in the dataset generator will account for the more large scale intereference that will be experienced \gls{ota}.

\subsection{Dataset Generation} \label{sec:generator}

\begin{figure}[t!]
    \centering
    \includegraphics[width=\columnwidth]{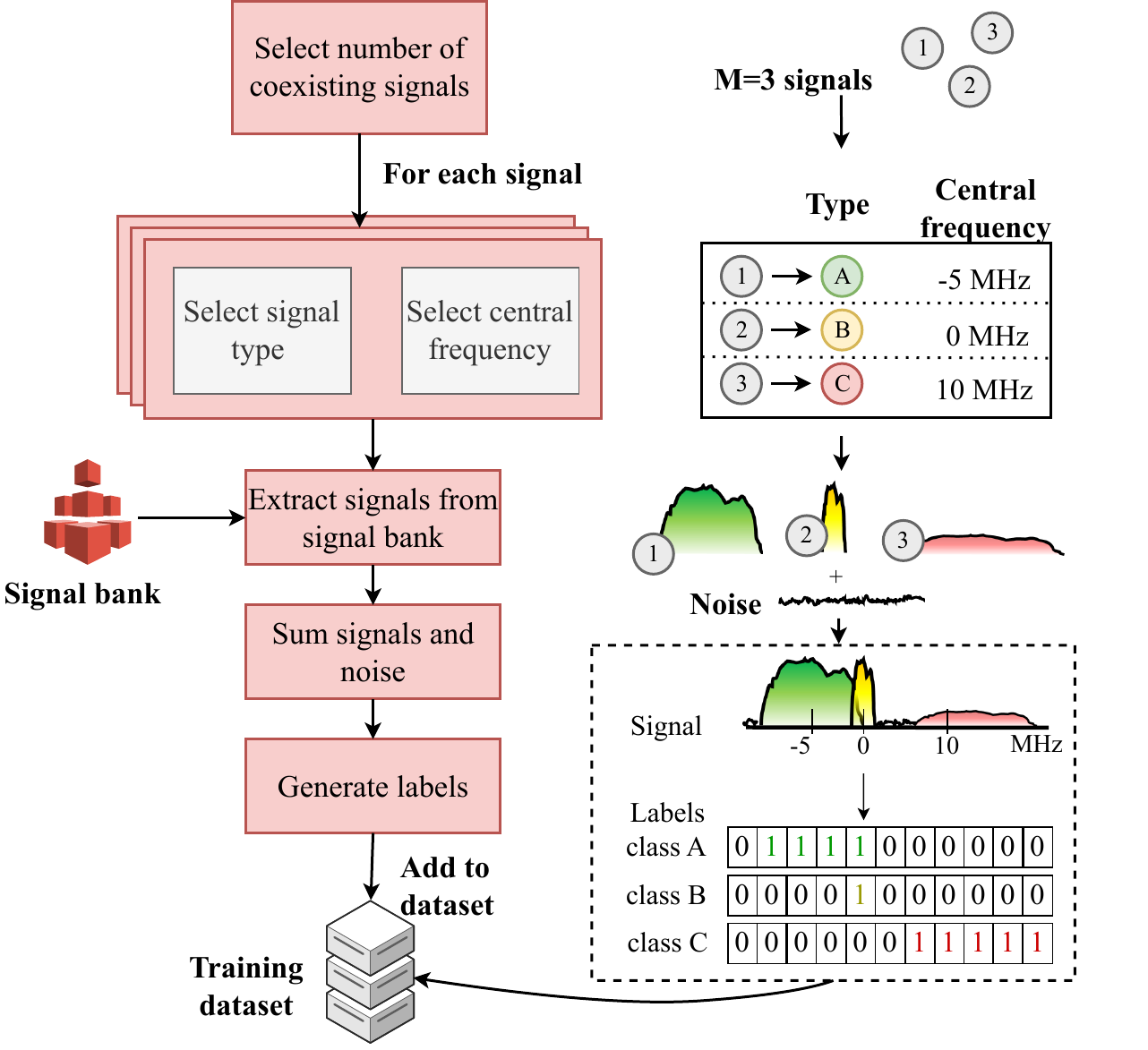}
    \caption{Generating a single training sample for the dataset.} 
    \vspace{-0.5cm}
    \label{fig:data_gen} 
\end{figure}

The next step of the semi-augmented dataset generator pipeline is combining multiple signals to generate a ``stitched'' wideband signal to be added to the training dataset. The steps are outlined in Fig. \ref{fig:data_gen} (left), while Fig. \ref{fig:data_gen} (right) shows an illustrative example of a sample and its corresponding label. 

The Dataset Generator is given a set of parameters such as the total number $C$ of signal types (i.e., WiFi, LTE, BLE, among others) that are present in the signal bank; the desired observable bandwidth $B$ (i.e., the field of view) of the receiver; the maximum number $n_s$ of signals which can be simultaneously present in $B$ at a given time; the probability $p_e$ that the entire observable bandwidth is empty; and the probability $p_c$ that any one of the signals is located at the center frequency. Similarly to real-world spectrum, our pipeline allows signals to overlap both partially or completely with other signals.


For each composite signal, we first generate the number $M\leq n_s$ of signals that will be injected into the observable bandwidth. This number is generated at random as follows:
\begin{equation}
M = \label{eq:m}
\begin{cases}
0, & p_e\\
U^{\mathrm{int}}(1,n_s), & 1-p_e
\end{cases}
\end{equation}
\noindent
where $U^{\mathrm{int}}(a,b)$ represents the value of a uniformly distributed integer random variable taking values in the range $[a,b]$. From \eqref{eq:m}, the band is completely empty with probability $p_e$, or it contains a number $M$ of signals uniformly chosen between $1$ and $n_s$. 
The probability $p_e$ makes it possible to generate examples that mimic diverse traffic conditions. To emulate a congested spectrum like the one in the ISM band, $p_e$ should be low so as to reflect the fact that such bands are frequently used by several devices at the same time. Similarly, less crowded scenarios can be emulated using a value of $p_e$ closer to $1$.

Upon randomly generating the number $M$ of signals to be included in the current example, the pipeline proceeds in assigning a target class to each of them. Specifically, let $\mathcal{C}=0,1,\dots,C$ be the set of $C+1$ possible class labels (e.g., where the class $0$ is reserved for empty portions of the spectrum and is not included in the signal bank), the target class $c_m$ of the $m$-th signal is randomly generated according to $c_m\sim U^{\mathrm{int}}(1,C)$. Note that $c_m$ does not include the class ``0" as unused portions of the spectrum will be computed in the last step of the pipeline. Upon determining the class $c_m\in\mathcal{C}\setminus\{0\}$ of signal $m$, the next step consists in extracting at random one instance from the signal bank and placing it in the spectrum. The positioning of the signal strongly depends on its bandwidth $b_m$, and its center frequency $f_m$. $f_m$ is randomly chosen so as to ensure that at least a portion of the signal appears within the observable band $[-B/2,B/2]$. Specifically,
\begin{equation}
f_m=
\begin{cases}
0, & p_c\\
U^{\mathrm{cont}}(-\frac{B}{2}-\frac{b_m}{2},\frac{B}{2}+\frac{b_m}{2}
), & 1-p_c
\end{cases}
\label{eq:center_freq}
\end{equation}
\noindent
where $U^{\mathrm{cont}}(a,b)$ represents the value of a uniformly distributed continuous random variable in the range $(a,b)$.

From \eqref{eq:center_freq}, the signal is centered at $0$~Hz with probability $p_c$, or it is centered at any frequency in $(-\frac{B}{2}-\frac{b_m}{2},\frac{B}{2}+\frac{b_m}{2})$ to ensure that the signal is at least partially present in the observable band. 

Once all of the $M$ signals have been generated and positioned in the spectrum, they are all combined together via an additive operation (or stitching). Finally, background noise measured \gls{ota} from an empty wireless channel is also added as  a "background" to fill empty spectrum portions. Upon generation of the sample, we produce its label (see the bottom right part of Fig. \ref{fig:data_gen}) and both are stored in the dataset. Since our goal is to produce a high-resolution and accurate labeling, labels were structured as a matrix $\mathbf{L}$ of dimension $C \times n_{iq}$, where $n_{iq}$ is the number of \glspl{iq} in frequency fed as input to the \gls{dl} model. For any given label, each row $i$ corresponds to a class in $\mathcal{C}$, and each column $j$ corresponds to a sub-band $k=1,2,\dots,n_{iq}$ of the observable bandwidth. Therefore, when a class $i$ is present in a specific sub-band $k$, the generic element $l_{i,j}$ of the matrix $\mathbf{L}$ is such that $l_{i,j}=1$ if a signal of class $i$ is present in sub-band $j$, $l_{i,j}=0$ otherwise. Thus, the resolution of our classification (e.g., how small of a sub-band we are able to classify) is therefore determined by $n_{iq}$. For example, if $B=25MHz$ and $n_{iq}=1024$ then we break the observable band into $n_{iq}$ bins of size $F=B/n_{iq}\approx24kHz$. 

\section{Semantic Spectrum Segmentation} \label{sec:seg}

In this section, we showcase one particular application of practical relevance where the proposed semi-augmented data generation pipeline can effectively enable the training of \gls{dl}-based solutions for spectrum sensing tasks that are capable of generalizing and delivering high accuracy and reliability. Specifically, we focus on the case of wide-band \textit{multi-label, multi-class spectrum sensing} where the goal is to accurately detect, characterize and localize multiple wireless signals with possibly diverse waveforms, bandwidths, power levels, and center frequencies but all coexisting (and possibly overlapping) within the same band of interest. 


Since multi-label, multi-class spectrum sensing aims at producing an accurate report of spectrum utilization, in practical applications it can be used to identify which portions of the spectrum are being used by a specific technology (e.g., WiFi, Bluetooth, and/or incumbents), while at the same time detect spectrum holes for opportunistic spectrum access in real-time. 


\subsection{Proposed Approach} \label{sec:proposed_semantic}

In contrast to conventional wideband spectrum sensing, we pursue a different approach and apply \textit{semantic segmentation}~\cite{ronneberger2015u}. This approach is well-established in the computer vision community as an effective tool to determine the shape of different elements in images or video frames. If compared to \gls{yolo}, which outputs a rectangular box around a specific object, semantic segmentation offers a much higher resolution by identifying all pixels that represent the object, therefore better conforming to the object's shape.


\begin{figure}[h]
    \centering
    \includegraphics[width=\columnwidth]{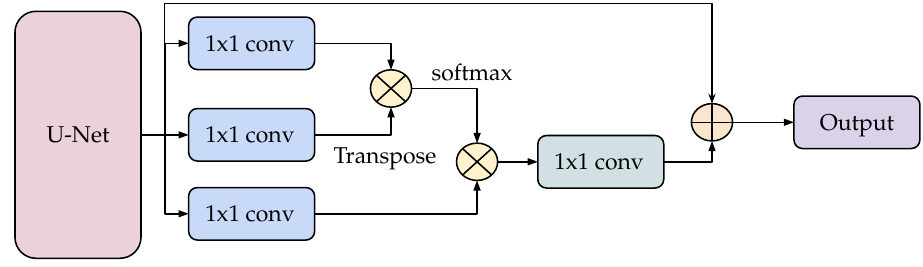}
    \caption{Proposed Semantic Spectrum Segmentation architecture utilizing U-Net \cite{ronneberger2015u} with a non-local block} 
    \label{fig:non-local} 
\end{figure}

In the context of spectrum sensing, semantic segmentation allows to detect multiple signals simultaneously and in one stage. Among other advantages, semantic segmentation is implemented as a single \gls{dl} classifier that can be optimized end-to-end by feeding \gls{iq} samples in the frequency domain as input.



The semantic spectrum segmentation model architecture is shown in Fig. \ref{fig:non-local}. Our architecture is inspired by U-Net~\cite{ronneberger2015u} to handle the majority of the feature extraction. To feed \gls{iq} in the frequency domain to the network, we adapt U-Net by converting 2D \gls{cnn} kernels to 1D. Our U-Net-based semantic spectrum sensing network consists of five encoding and decoding blocks. The encoding block is a stack of two 1D convolutional layers with a kernel size $1\times3$ followed by batch normalization and ReLU activation. A maxpooling layer is used to downsample the encoded features. The decoding block takes both output features from its previous layer and the encoded features from a skip connection. To match the dimensionality of both inputs, an upsampling layer followed by a convolutional kernel is applied to the downsampled features. Feature dimensions are 64, 128, 256, 512, and 1024 for the five encoding and decoding blocks, respectively. 


\subsubsection{\textbf{Non-local Block}} One of the issues with \glspl{cnn} is that they tend to process groups of features locally, which makes it difficult to capture spatially distant information without the use of feature merging techniques such as the use of pooling layers. Since U-Net has neither global pooling layers nor fully connected layers to mix up features globally, utilizing U-Net in its original form results in performance loss, especially when dealing with wideband signals in the presence of interference. To improve the system performance and increase accuracy, we apply a Non-local block \cite{wang2018non} after the final decoding block of U-Net so as to integrate a self-attention process. This is achieved via the architecture shown after U-Net in Figure \ref{fig:non-local}. Specifically, three $1\times1$ convolutional kernel are used to encode input as Queries $\mathbf{Q}$, Keys $\mathbf{K}$, and Values $\mathbf{V}$, respectively. The self-attention is instead defined as:
\begin{equation}
    \mathrm{Attention}(\mathbf{Q},\mathbf{K},\mathbf{V})=\mathrm{softmax}\left( \frac{\mathbf{Q}\mathbf{K}^T}{\sqrt{d}} \right) \mathbf{V}
\end{equation}
\noindent
where $\sqrt{d}$ is the embedding dimension of $\mathbf{Q}$ and $\mathbf{K}$. The self-attention computes a weighted average of the whole encoded input features $\mathbf{V}$ based on an attention map $\mathrm{softmax}(\mathbf{Q}\mathbf{K}^T/\sqrt{d})$. 

Such an attention technique can apply to every encoding and decoding block in U-Net to mix up features globally and improve performance. However, the non-local block requires two matrix multiplication over the whole spatial and feature dimension, which is a relatively computationally heavy task. This complexity might result in long inference times that might make the sensing output obsolete by the time it is computed. Since our goal is to develop a system that can operate in real-time and in an actual wireless deployment, we have decided to add a non-local block between the last decoding block and the final multi-labeling layer only.


\subsubsection{\textbf{Extension to multi-label semantic segmentation}} One of the main aspects that make semantic segmentation for spectrum sensing applications significantly different from its traditional application to computer vision tasks is that objects in the foreground always hide those in the background, but the same does not hold for \gls{rf} applications. Pixels in a picture only carry information about foreground objects. However, two wireless signals can overlap in both time and frequency domains and coexist at the same time without necessarily being hidden by other signals in the same frequency bands. The conventional semantic segmentation approach can only assign a single class to each pixel (e.g., the class of the object in the foreground or the signal with the highest power), therefore it cannot be used directly to perform spectrum sensing tasks as it would necessarily result in misclassifications when applied to overlapping wireless signals

To solve this problem, we have extended the traditional semantic segmentation architecture to consider the much more complex task of multi-label multi-class classification where the output of the system is a multi-dimensional binary mapping between each \gls{iq} sample and all possible classes. More specifically, we have extended the U-Net architecture such that the input is represented by a sequence of \gls{iq} samples in the frequency domain with size $2\times n_{iq}$, where $n_{iq}$ denotes the number of frequency bins or \glspl{iq} (i.e., the resolution of the classifier), and 2 denotes the real and imaginary part of the complex-values \gls{iq} samples. A $1\times1$ convolutional kernel at the last layer will generate a $C\times n_{iq}$ output, where $C$ denotes the number of classes. Thus, each row of the output is a binary segmentation map where, for each \gls{iq} sample, we can extract information on which classes are present or absent in the specific frequency bin. \smallskip



\subsubsection{\textbf{Achieving Generalization}}~While the dataset generator helps generalize across different bandwidths, center frequencies, and wireless protocols, noise estimation helps generalize across different wireless devices or radios. Specifically, during training, we record the minimum values of the smoothed signal power in the frequency domain of each sample and use the average of those values as a reference estimate of the noise floor. When testing the model, we multiply the input signal by a factor that, similarly to normalization, is computed to shift the minimum signal power to the same level as that we have estimated across the training dataset. However, such an estimate assumes that there will be an empty hole in the spectrum that only contains noise (this is very common in wideband spectrum sensing scenarios). A more sophisticated noise estimation approach may help to further improve the generalization.
\begin{figure}[h!]
    \centering
    \includegraphics[width=0.98\columnwidth]{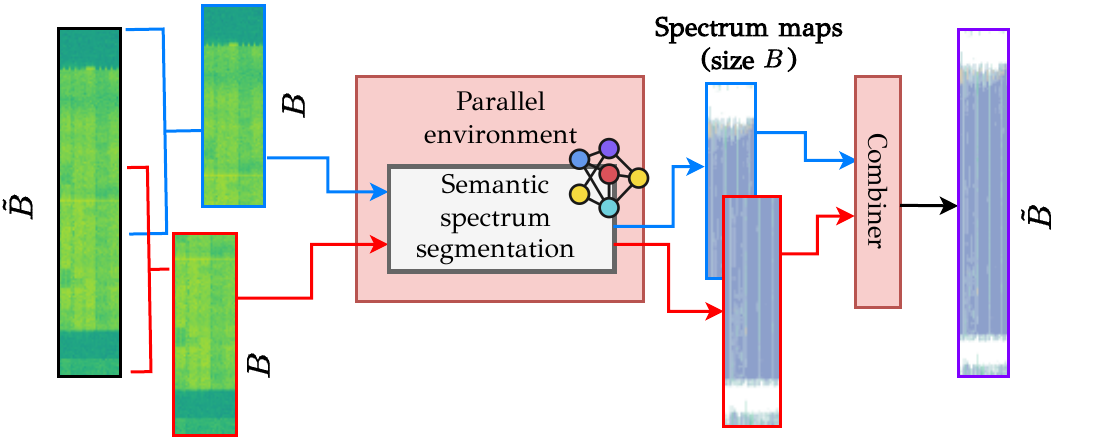}
    \caption{Scalable and portable pipeline for processing larger bandwidths without retraining the segmentation classifier.} 
    \label{fig:parallel}
\end{figure}

\subsubsection{\textbf{Scalable and Portable Wide-band Processing}}~As mentioned in previous sections, the resolution of our approach is $F=\sfrac{B}{n_{iq}}$ and depends on both the size of the observable bandwidth $B$ and a number of \gls{iq} samples in the frequency domain (e.g., the number of sub-bands used to split the bandwidth $B$). Since the semantic spectrum segmentation classifier is trained to process an observable band $B$, it can indeed process signals with smaller bandwidths, but cannot process those with a larger bandwidth due to a mismatch between the size of the input signal, and that of the classifier. To support increasingly larger bandwidths, there are two approaches. A naive one would be to generate a new dataset with a larger observable bandwidth and train a new classifier. A more scalable and portable approach, which is the one we follow in our architecture, is instead to leverage parallel processing (e.g., GPU parallelism) and developed a pipeline that can be adapted to any observable bandwidth without requiring to retrain the classifier. The pipeline is illustrated in Fig. \ref{fig:parallel}. Whenever we receive an input covering a portion of spectrum $\tilde{B}$ that is larger than the observable bandwidth $B$ of the classifier, we first collect $\tilde{n}_{iq} = \sfrac{\tilde{B}}{B} \cdot n_{iq}$ samples to process. We then generate a set $N$ of partially overlapping samples from the original input such that each sample covers only a portion of size $B$ of the input signal. Each individual sample is then processed individually by our classifier to obtain $N$ outputs of size $n_{iq}$. These are then combined together via averaging to produce a final output of size $\tilde{n}_{iq}$. The advantage of this pipeline is that (i) it is scalable in that it can leverage parallel processing to speed up the inference time for inputs covering larger bandwidths than $B$; and (ii) it is portable and general as it can be used for virtually any input of variable bandwidth $\tilde{B}$, thus making it portable. 

\section{Performance Evaluation} \label{sec:results}

\subsection{Generating the experimental dataset}

To demonstrate the effectiveness of our approach, 
our signal bank is generated by collecting data for three hours across three days and using only two radios, a USRP X310 (i.e., the transmitter) and a USRP N320 (i.e., the receiver). Data is collected using five classes ($C=5$) where each class corresponds to a wireless protocol/technology: WiFi, LTE, BLE, LoRa, ZigBee. The spectrum sensing receiver utilizes a sampling rate of 25MHz which is also our chosen observable bandwidth ($B=25MHz$). We set $n_s=2$ and $p_e=0.05$ to mimic the case of a congested ISM band where the chances of having an empty channel are very small. Finally, we set $p_c=0.5$ as we aim at emulating receivers tuned on commonly used center frequencies (e.g., $2.437$~GHz for WiFi channel 6, $2.655$~GHz for LTE band 7). 

For our data collection, we consider a laboratory area that hosts Arena~\cite{bertizzoloComnet20}, a 64-antenna indoor testbed with ceiling-mounted antennas distributed according to an 8$\times$8 grid covering a 2240 square feet office space in our campus. In Arena, antennas are connected to USRP X310 and N320 radios which, together with the ceiling-mounted antennas, are used to collect the \gls{ota} data. We also consider two other areas open to the public, i.e., a kitchen area and the main lobby of a building that hosts classes, a cafeteria, and several study areas. These three areas are characterized by diverse RF channel conditions. 

\begin{figure}[t!]
    \centering
    \includegraphics[width=0.95\columnwidth]{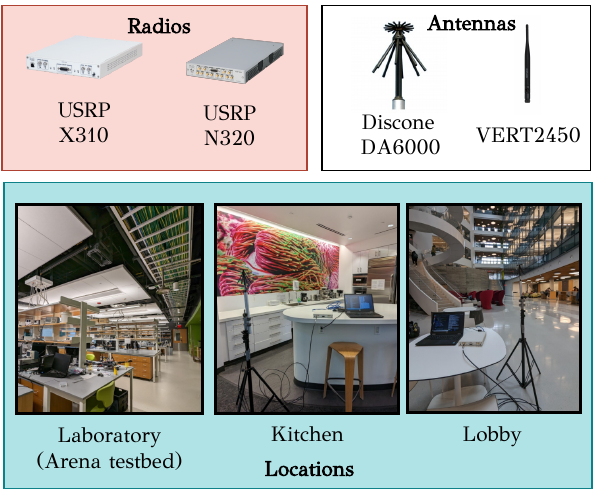}
    \caption{Setup used for data collection with different radios, antennas, and in different locations. The signal bank has been generated in the laboratory location only, while testing data has been collected in all locations.}
    \vspace{-0.4cm}
    \label{fig:setup}
\end{figure}

The hardware setup used to collect the data and the data collection locations are illustrated in Fig. \ref{fig:setup}.
The data used to generate our signal bank has been collected in the laboratory setup only, while the data we use in this section to evaluate our solution has been collected in all of the three locations. For the signal bank, we collect data around the 900 MHz, 2.4GHz, and 2.6GHz bands. We find 25MHz wide bands with minimal interference and collect data while monitoring the spectrum to remove data collected in the presence of strong interference. 

\begin{figure}[h!]
    \centering
    \includegraphics[width=0.9\columnwidth]{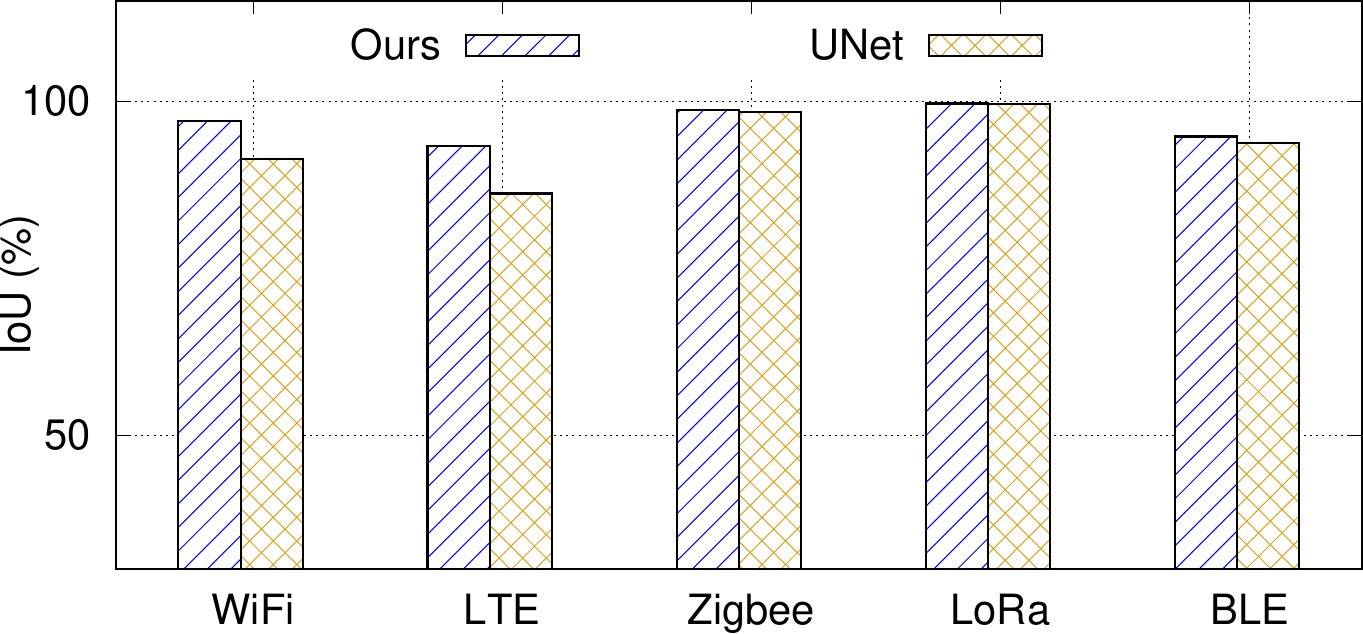} 
    \caption{\gls{iou} comparison for different models and wireless technologies.}
    \vspace{-0.5cm}
    \label{fig:baseline}
\end{figure}



\subsection{Classification Performance} \label{sec:accuracy_res}

To evaluate the performance of our semantic segmentation model, we compare it with the original U-Net~\cite{ronneberger2015u}. The mean \gls{iou} are 96.70\%, 93.44\% for ours and U-Net respectively. Figure~\ref{fig:baseline} demonstrates the \gls{iou} performance for each wireless technology, in which ours can achieve 97.08\%, 93.29\%, 98.72\%, 99.65\% and 94.74\% for WiFi, LTE, Zigbee, LoRa and BLE signals. On the other hand, the U-Net has 90.31\%, 85.20\%, 98.40\%, 99.59\% and 93.72\% \gls{iou} for five technologies respectively. 
\begin{figure}[t!]
    \centering
    \includegraphics[width=0.98\columnwidth]{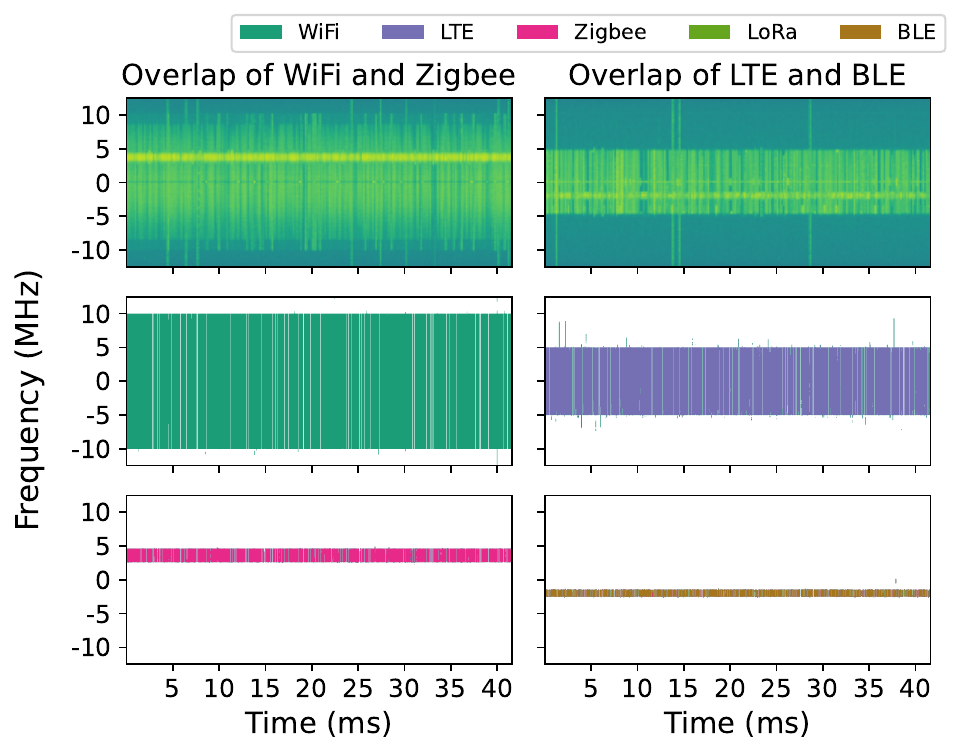}
    \caption{Two examples of classification for overlapping signals: (Left) WiFi and Zigbee; (Right) LTE and BLE.}
    \vspace{-0.5cm}
    \label{fig:overlap}
\end{figure}
Although our model is 3\% better than U-Net in general, it can achieve 7\% accuracy better than U-Net on signals with larger bandwidth such as WiFi and LTE due to the non-local block that we have introduced exactly to improve the detection of signals that occupy large portions of spectrum.
\begin{figure}[b]
    \centering
      \includegraphics[width=\columnwidth]{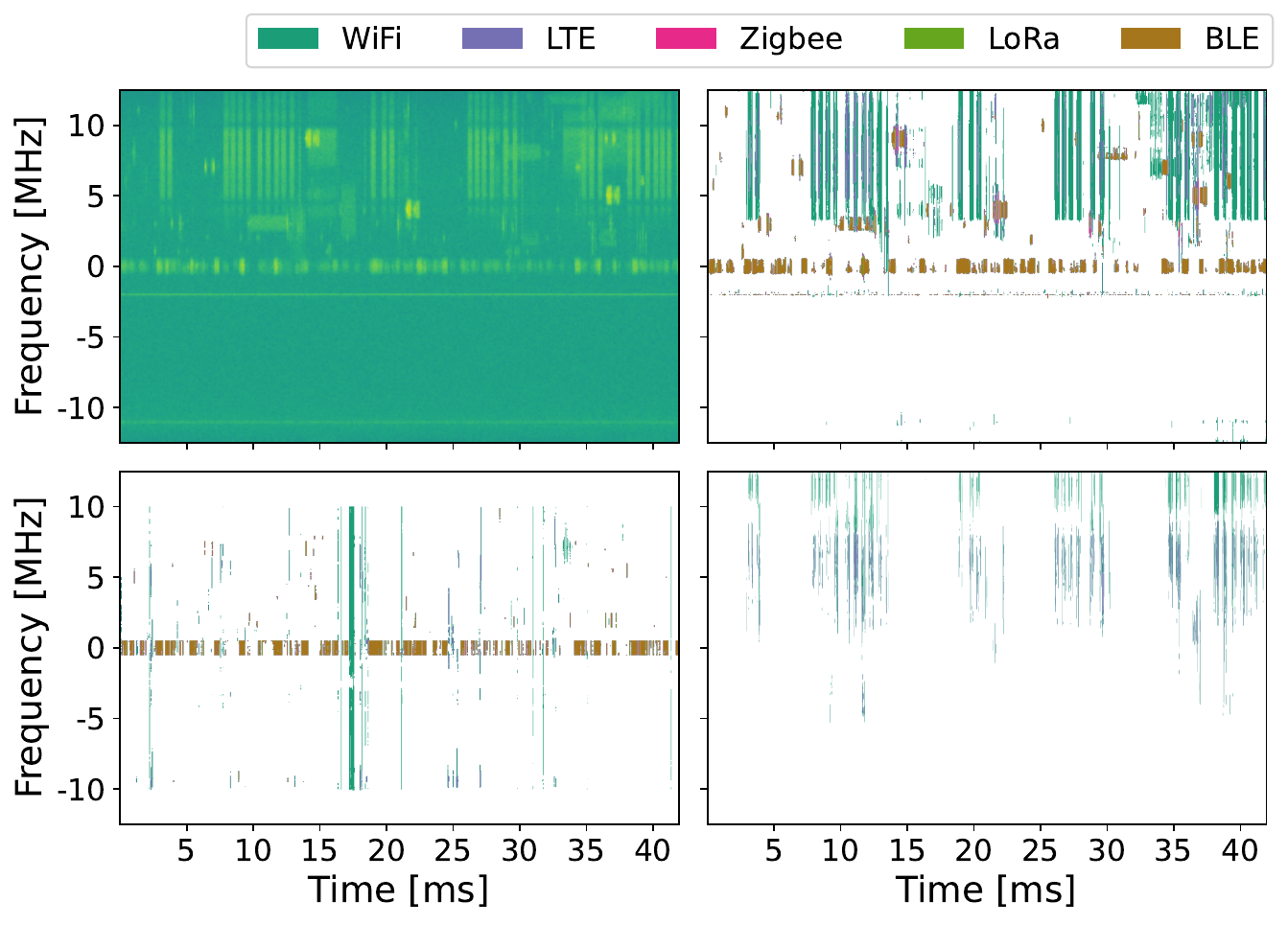}
    \caption{Spectrum maps of our classifier trained on three different datasets and tested on data collected \gls{ota}: (Top Left) Spectrogram of data collected at 2.402 GHz (i.e., BLE advertisement channel); (Top Right): Proposed semi-augmented data generator model output; (Bottom Left):  Un-augmented \gls{ota} data model output; (Bottom Right):  Synthetic data model output.}
    \label{fig:augment}
\end{figure}

\begin{figure*}[t]
    \centering
    \includegraphics[width=0.8\textwidth]{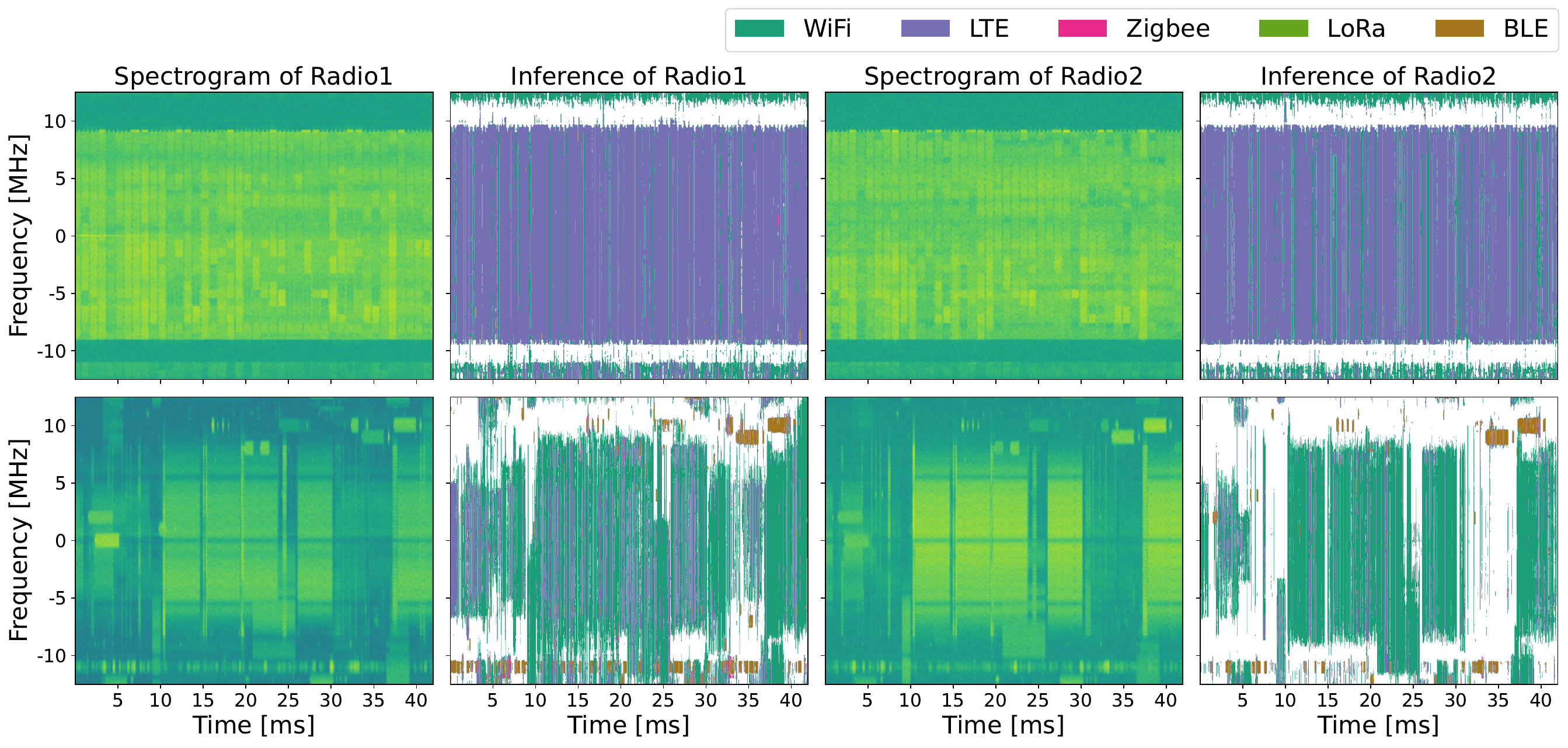}
    \caption{Spectrum maps for data collected \gls{ota} with different radios and antennas: (Top) Data collected at 748 MHz (i.e., around LTE band 28); (Bottom): Data collected at 2.437 GHz (i.e., WiFi Channel 6).}
    \label{fig:ant_results}
    \vspace{-0.5cm}
\end{figure*}

 As mentioned in Section \ref{sec:proposed_semantic}, our semantic spectrum segmentation classifier outputs a binary array of size $C\times n_{iq}$ to determine which type of signal is present in each \gls{iq} sample. In Fig. \ref{fig:overlap}, we show two examples of inputs and outputs to illustrate the ability of our solution in detecting and individually classifying with high-resolution overlapping signals of different technologies. We show the class-specific output for the classes related to the overlapping signals only. Fig. \ref{fig:overlap} (left) shows how the classifier is able to effectively classify simultaneously overlapping WiFi and Zigbee signals, while Fig. \ref{fig:overlap} (right) shows the same but for LTE and BLE overlapping signals.

\subsection{Experimental Validation In the Wild } \label{sec:experiment}
\vspace{-0.1cm}


\begin{figure}[t]
    \centering
    \includegraphics[width=\columnwidth]{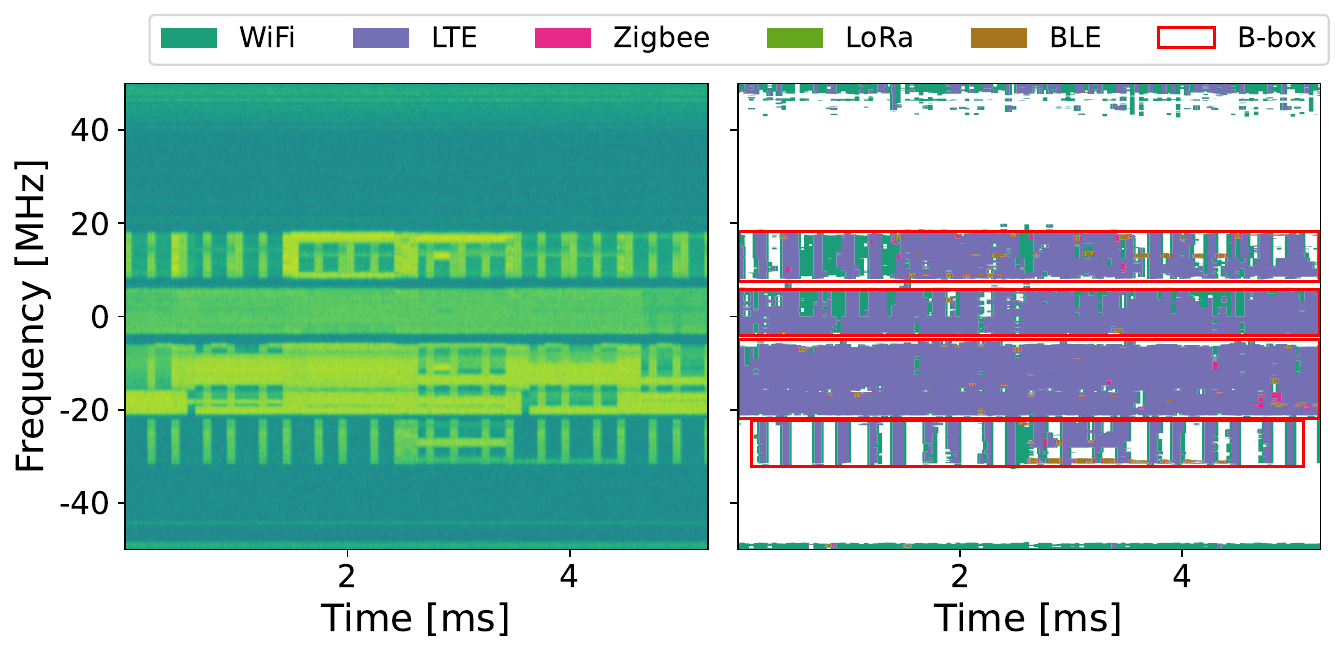}
    \caption{(Left) Spectrogram of \gls{ota} data  at 100 MHz sampling rate in 700-800MHz. (Right) Our model output with ideal object detection bounding box.} 
    \vspace{-0.5cm}
    \label{fig:samp_results}
\end{figure}

Our main goal is to attain reliable and effective spectrum sensing in practical use cases \gls{ota} and \gls{itw}. For this reason, we have performed an extensive data collection campaign, \textit{purely for testing not for training}. This data spans multiple days at different locations where we have collected a variety of \gls{ota} signals being transmitted at different center frequencies, and with different sampling rates, radios, and antennas. The goal of this campaign is to collect as much data as possible to validate our solution in a variety of real-world deployments, as well as to showcase the benefits that our system design brings to the final accuracy of the spectrum sensing task.


Since the signals we have collected in this case are collected purely \gls{itw} in a completely uncontrolled RF environment shared by students, workers, visitors, and other wireless devices,
it is challenging (if not impossible) to build an accurate and comprehensive ground truth for signals 
Therefore, we follow the approach used in computer vision problems with unlabeled test data and resort to graphical comparison of the outputs only.

\subsubsection{\textbf{Performance with Varying Training Datasets}}~In Fig. \ref{fig:augment} 
we show the output of our classifier for \gls{ota} spectrum measurements when the classifier is trained using three different datasets: an augmented dataset obtained using the proposed semi-augmented dataset generation pipeline described in Section \ref{sec:generator}, the un-augmented \gls{ota} dataset used to generate the former, and a synthetic one generated in MATLAB and without any \gls{ota} data, containing artificial channel effects. 


Signals are collected on the ISM band when the receiver center frequency is on a BLE advertisement channel (2.402 GHz).  We see that the classifier trained on data produced by our augmented dataset generator is able to classify and localize the centered BLE signals better than the same classifiers trained using both the un-augmented and synthetic datasets. Moreover, we notice that the classifier trained over the augmented data is able to classify and localize signals not necessarily centered at the center frequency of the receiver. 
The classifier trained on the un-augmented dataset can classify the BLE signal to some extent but misclassifies WiFi signals even in those areas of the spectrum where there is no activity. The classifier trained on the synthetic dataset, instead, is completely unable to classify any of the \gls{ota} signals since synthetic data does not properly capture the characteristics of wireless signals collected \gls{itw}.


\subsubsection{\textbf{Performance with Varying Radios}}~Fig. \ref{fig:ant_results} shows the performance of the classifier on testing data collected from two different radios. Radio 1 being the USRP N320 equipped with the DA6000 antenna (which is also the same setup we used to collect the data used to generate the augmented dataset), and Radio 2 being the USRP X310 with a VERT2450 antenna. Similarly to the previous case, each row corresponds to an individual signal collected at different center frequencies and bands. In this case, the top figure shows the spectrogram of a signal collected in the LTE band at 748 MHz, while the bottom one has the same center frequency as WiFi channel 6. Indeed, we can see that the output of the classifier for both radios is very similar and show good classification performance in both cases, but with Radio 1 (i.e., the USRP N320) detecting more signals than Radio 2 (i.e., the USRP X310). For example, Radio 1 produces \gls{iq} samples that make it easier to detect the BLE signals, while Radio 2 can detect them only partially.


\subsubsection{\textbf{Performance with Varying Sampling Rates}}~We also evaluate the generalization capability of the classifier by testing it on bandwidths larger than the ones used during the training process. Specifically, we utilize data sampled at 100MHz centered at 750MHz, a known LTE band.
We broke up the larger bands into overlapping pieces and averaged the classifications of the overlapping segments. We decided to take this approach to show that you do not necessarily need to retrain the network for higher bandwidths and can either divide and process pieces in parallel if you have the higher sampling capability or even sweep across large bands to cover them if you do not. 
Fig. \ref{fig:samp_results} shows the model output when tested on this data. We see there is still good localization and classification performance even as we process these higher sampling rates. The model can accurately detect and localize the LTE signals which are common in this range of frequencies. Furthermore, we also show a set of red bounding boxes that reflect a perfect object detection output from YOLO. It is clear to see that rectangular bounding boxes do well in locating LTE transmissions, but mark spectrum holes as LTE signals, whereas our segmentation outputs are more accurate in detecting such holes.


\subsubsection{\textbf{Performance with Varying Locations}}
\begin{figure}[t]
    \centering
    \includegraphics[width=\columnwidth]{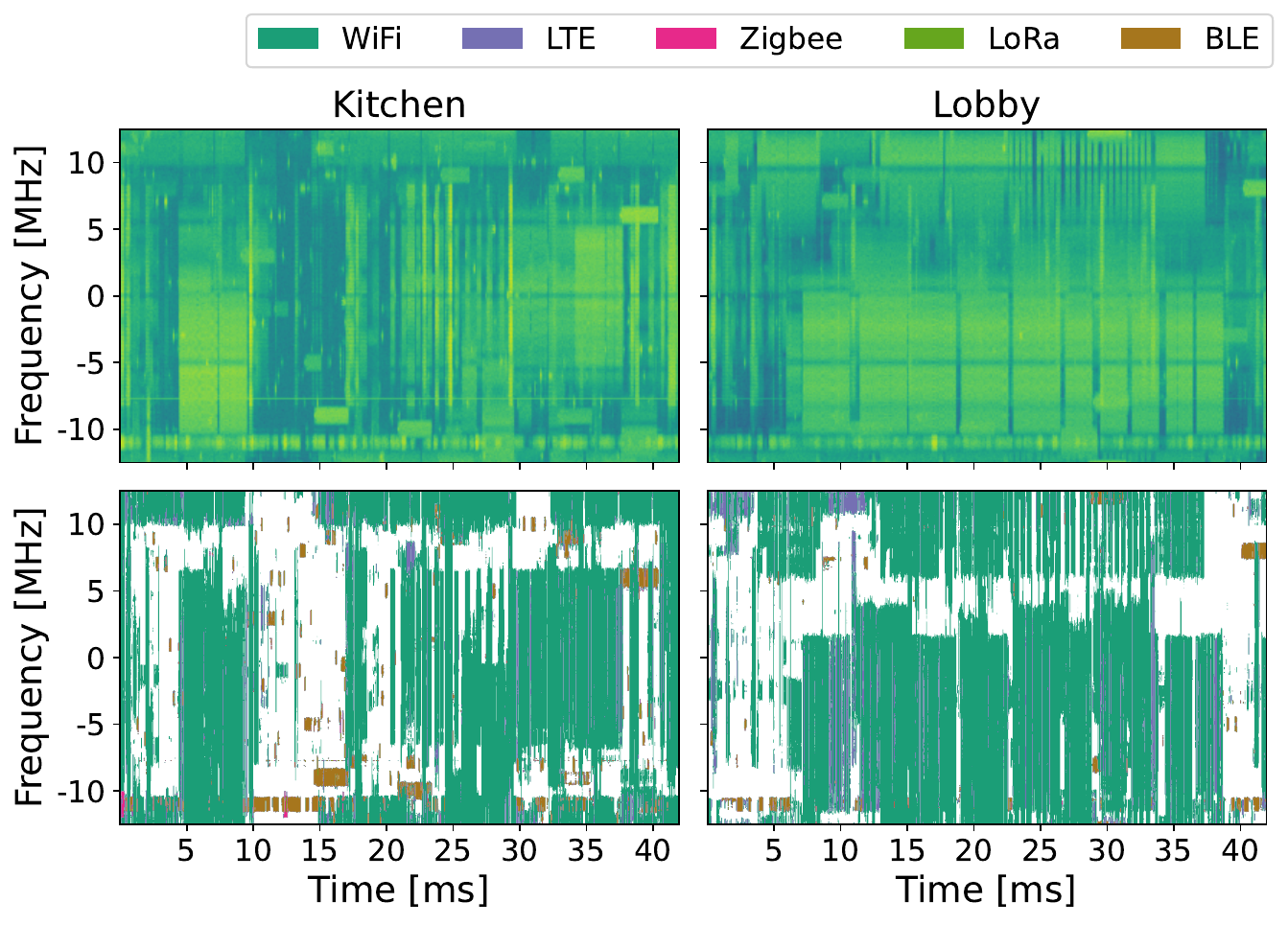}
    \caption{Spectrum maps for data collected \gls{ota} at different locations with $B=25$~MHz and at $2.437$~GHz: (Left) kitchen area; (Bottom): main lobby.}
    \vspace{-0.5cm}
    \label{fig:diff_location}
\end{figure}

~Fig. \ref{fig:diff_location}
shows our model performance in two other physical locations at 2.437 GHz (i.e., WiFi channel 6). The two columns show the spectrogram and output of the kitchen and lobby. In both cases, the model accurately detects the WiFi signals dominating this center frequency and is able to also detect the BLE signals off to the side of the band. Therefore our model is resilient to changes in physical location as well.






\subsection{Evaluation of Inference Time}

\begin{figure}[!t]
    \centering
\includegraphics[width=0.95\columnwidth]{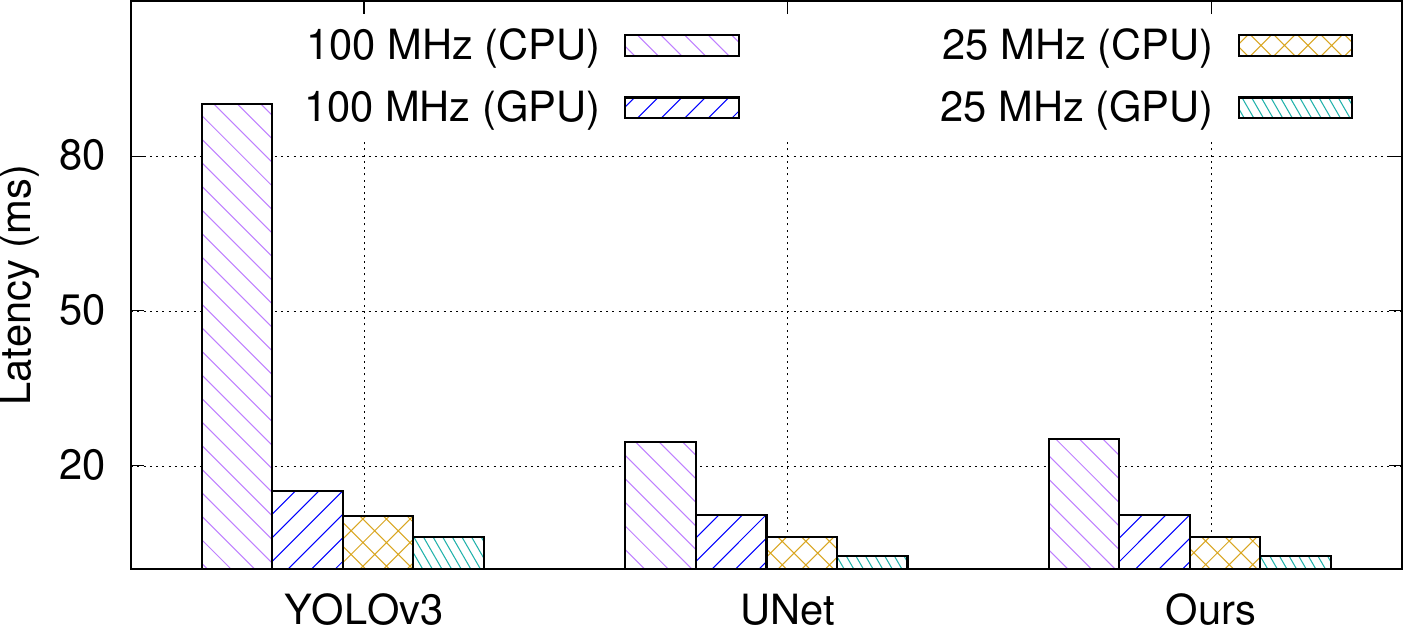}
    \caption{Inference latency comparison on CPU and GPU for different observable bandwidth size values.}
    \label{fig:comp_lat_model}
    \vspace{-0.5cm}
\end{figure}

To assess the suitability of our model to deliver real-time spectrum sensing capabilities, we evaluate its inference time on both CPU and GPU and compare it with YOLOv3 (as used in \cite{nguyen2023wrist}) and the original U-Net model. Our results have been obtained by averaging 1000 independent measurements. 
Fig.~\ref{fig:comp_lat_model} shows the latency tested on different bandwidths. Because YOLOv3 is generally used on 2D images, we assume an input size of 512x512 for 100MHz, as done in \cite{nguyen2023wrist}, and scale it down linearly to 128x128 for 25MHz. As expected, the YOLO model has the highest latency since it deals with larger input sizes and requires more complex computations. We only report the latency to compute an output from YOLOv3 without including the time needed to perform non-maximum suppression, i.e., the final step of object detection tasks where the most appropriate bounding box is chosen (which would further increase latency). The difference in latency between our model and YOLOv3 is substantial, with our model being able to compute an output up to 31\% faster than YOLOv3 when running on a GPU at 100~MHz. However, when the models are executed on CPUs, our model is 72\% faster then YOLOv3. On multiple ends, our model significantly outperforms YOLOv3 by a wide margin both on a CPU or GPU.
Interestingly, although our model extends U-Net by adding a non-local block and a multi-label segmenter, its inference time is almost identical to that of the original U-Net model. Specifically, if compared to U-Net, our model is 0.126~ms slower on CPU \textit{(2\% increase)}, and 0.007~ms slower on a GPU \textit{(0.25\% increase)}. 

\section{Conclusions and Remarks} \label{sec:conclusion}

In this work, we have proposed a novel approach based on \textit{semantic spectrum segmentation}. Moreover, we have proposed a novel data generation approach where a limited set of easy-to-collect real-world wireless signals are ``stitched together'' to generate large-scale, wideband, and diverse datasets. We have extensively evaluated our approach through experiments on multiple testbeds (including the Arena testbed) using multiple antennas, multiple sampling frequencies, and multiple radios over the course of 3 days. Our results have shown that our approach classifies and localizes signals with a mean intersection over union (IOU) of 96.70\% across 5 wireless protocols while performing in real-time with a latency of 2.6 ms. Moreover, we have demonstrated that our approach based on non-local blocks achieves 7\% more accuracy when segmenting the most challenging signals with respect to the state-of-the-art U-Net algorithm. Public access to our code and data is available at \url{https://github.com/uvaydovd/spectrum\_sensing\_stitching}.

\section*{Acknowledgment of Support and Disclaimer}

This work is funded in part by the National Science Foundation grants CNS-2134973, CNS-2120447, ECCS-2229472, ECCS-2146754, CNS-2112471 and ECCS-2329013,  and in part by funds from OUSD R\&E, NIST, and industry partners as specified in the Resilient \& Intelligent NextG Systems (RINGS) program and in the Future of Semiconductors (FuSe) program, by the Air Force Office of Scientific Research (AFOSR) under contract number FA9550-23-1-0261, by the Office of Naval Research (ONR) under award number N00014-23-1-2221, and by the Office of the Director of National Intelligence (ODNI), Intelligence Advanced Research Projects Activity (IARPA), via [2021-2106240007]. The views and conclusions contained
herein are those of the authors and should not be interpreted as
necessarily representing the official policies, either expressed
or implied, of NSF, AFOSR, ONR, ODNI, IARPA, or the U.S. Government. The
U.S. Government is authorized to reproduce and distribute reprints for governmental purposes notwithstanding any copyright annotation therein.

\balance
\footnotesize
\bibliographystyle{IEEEtran}
\bibliography{reference}

\end{document}